\documentclass[aps,prl,showpacs,preprint]{revtex4}

\usepackage{graphicx} 
\usepackage{amsmath}

\begin{document}

\title{Refractive Index Enhancement with Vanishing Absorption in an Atomic Vapor}
\author{N. A. Proite, B. E. Unks, J. T. Green, and D. D. Yavuz}
\affiliation{Department of Physics, 1150 University Avenue,
University of Wisconsin at Madison, Madison, WI, 53706}
\date{\today}
\begin{abstract}
We report a proof-of-principle experiment where the refractive
index of an atomic vapor is enhanced while maintaining vanishing
absorption of the beam. The key idea is to drive alkali atoms in a
vapor with appropriate control lasers and induce a gain resonance
and an absorption resonance for a probe beam in a two-photon Raman
configuration. The strength and the position of these two
resonances can be manipulated by changing the parameters of the
control lasers. By using the interference between these two
resonances, we obtain an enhanced refractive index without an
increase in the absorption.
\end{abstract}
 \maketitle

Since the birth of quantum and nonlinear optics, one of the key
challenges has been if one can achieve a very large refractive
index for a laser beam \cite{ScullyBook}. A key application of a
large refractive index is to optical imaging science. It is
well-known that the wavelength of light inside a refractive medium
is $\lambda=\lambda_0/n$, where $\lambda_0$ is the wavelength in
free space and $n$ is the refractive index. A large refractive
index, therefore, corresponds to a reduced wavelength inside the
medium and enhanced imaging resolution. Another important
application of a large refractive index is to optical lithography
where the smallest feature size of a lithographic mask is
determined by the wavelength of light. A simple and an efficient
way to achieve a large refractive index is likely to have
significant practical implications since lithographic resolution
currently determines the size and the processing power of every
semiconductor integrated circuit.

A laser beam which is tuned close to an atomic resonance can
experience a large refractive index. As an example, the refractive
index for a gas at a pressure of 1 torr can reach values as high
as 10. However, such a large index is usually accompanied by large
absorption and  the effect is not useful. This is because, near an
optical resonance, the real and imaginary parts of the optical
susceptibility are of the same order.  It was first pointed out by
Scully that, by using interference in a three state atomic system,
it is possible to obtain a large refractive index with negligible
absorption \cite{Scully1,Scully2}. The pioneering work of Scully
was extended to different configurations by Fleischhauer and
colleagues \cite{Fleischhauer1,Rathe,Keitel}. Although these ideas
were experimentally demonstrated in a Rb vapor cell by Zibrov {\it
et al.} \cite{Zibrov}, it has not yet been possible to achieve a
refractive index in a vapor that is large enough to be of
practical importance. In this work, we report a proof-of-principle
experiment that demonstrates a new approach to this long-standing
problem \cite{Yavuz,Kocharovskaya1}. By utilizing the interference
of two Raman resonances, we show that the refractive index of a
laser beam that is very far detuned from an electronic resonance
can be enhanced while maintaining vanishing absorption.

Before proceeding with a detailed description of our experiment,
we note that in recent years, a number of counter-intuitive
effects in driven atomic systems have been predicted and
demonstrated
\cite{Harris1,Kocharovskaya2,Harris2,Hau1,Kash,Fleischhauer2,Phillips,Hau2,Wang,Boyd,Imamoglu,Harris3,Zhu}.
Of particular importance is the demonstration of slow light
\cite{Harris2,Hau1,Kash}, stopped light
\cite{Fleischhauer2,Phillips,Hau2}, fast and backward light
\cite{Wang,Boyd}, and enhanced nonlinearities using
Electromagnetically Induced Transparency
\cite{Imamoglu,Harris3,Zhu}. Broadly, most of these effects can be
thought as engineering the variation (the slope) of the refractive
index as a function of frequency. In contrast, our technique
allows us to engineer the actual value of the refractive index
while maintaining vanishing absorption to the beam. In this sense,
our approach complements the existing techniques in modifying the
optical response of an atomic medium.

In our experiment, we follow the suggestion of Yavuz {\it et al.}
\cite{Yavuz} which has recently been extended to high alkali
densities by Kocharovskaya and colleagues \cite{Kocharovskaya1}.
The essential features of this idea are presented in Fig.~1. It is
well-known that, the interference of the dipole moments of an
absorptive resonance and an amplifying resonance can lead to an
enhanced refractive index with vanishing absorption
\cite{Fleischhauer1}. As shown in Fig.~1(a), the most
straightforward way to realize such an interference would be to
have two different two-level atomic species. In practice, such a
multiple two-level atom scheme has not yet been realized since it
is difficult to find two different atomic species with very close
and easily tunable resonance frequencies. The key idea that we
experimentally demonstrate in this work is that such a multiple
two-level scheme can be realized by using Raman resonances in
far-off resonant atomic systems. As shown in Fig.~1(b), with an
atom starting in the ground state $|g\rangle$, a Raman transition
involves absorption of one photon and emission of another photon
of different frequency such that the two-photon resonance
condition is satisfied. By changing the order at which the probe
laser, $E_p$, is involved in the process, such a Raman resonance
can be made absorptive or amplifying. This approach circumvents
the difficulties of the scheme of Fig.~1(a). The two Raman
transition frequencies can be arbitrarily different since we have
the freedom to choose the frequencies of the control lasers,
$E_{c1}$ and $E_{c2}$.  Figure~1(c) shows the real and the
imaginary parts of the susceptibility for the case of equal
strength of the two resonances. The real part of the
susceptibility, $\chi'$, and therefore the refractive index, peaks
at the point of vanishing imaginary part, $\chi''$.

\begin{figure}[th]
\begin{center}
\includegraphics[width=13cm]{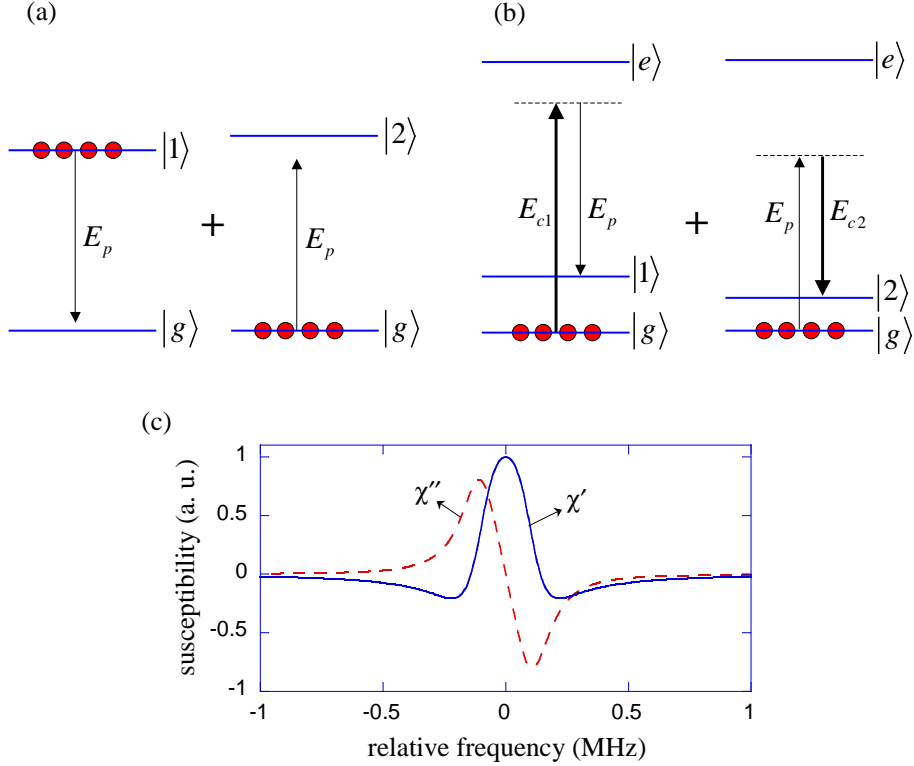}
\end{center}
\caption{(Color online) The interference of an absorptive
resonance and an amplifying resonance can lead to an enhanced
refractive index with vanishing absorption. (a) shows the most
straightforward way to achieve such an interference. Due to
various difficulties, the scheme in (a) is not practical (see text
for details). (b) shows an equivalent scheme that we
experimentally demonstrate in this work. With an atom starting in
the ground state $|g\rangle$, a Raman transition involves
absorption of one photon and emission of another photon of
different frequency such that the two-photon resonance condition
is satisfied. By changing the order at which the probe laser,
$E_p$, is involved in the process, such a Raman resonance can be
made absorptive or amplifying. (c) shows the real part, $\chi'$,
and the imaginary part of the susceptibility, $\chi''$, as a
function of frequency. Here, we take the two resonances to be of
equal strength with a width of 0.1~MHz and assume the spacing
between resonances to be 0.2~MHz.} \label{introduction}
\end{figure}

We demonstrate the scheme of Fig.~1(b) in two isotopes of atomic
Rubidium, $^{87}$Rb and $^{85}$Rb. Figure~2 shows the simplified
experimental set up and the relevant energy level diagrams. We
work with a triple layer magnetically shielded and temperature
controlled natural abundance Rb (28~\% $^{87}$Rb, 72~\% $^{85}$Rb)
vapor cell. The vapor cell is $L=7.5$~cm long and contains 10~torr
of Neon (Ne) as a buffer gas. The temperature of the vapor cell is
kept at $T=90$ degrees Celsius which gives a total atomic density
of $N \approx 2.4 \times 10^{12}$~/~cm$^3$. We use $F=2
\rightarrow F=1$ and $F=2 \rightarrow F=3$ hyperfine transitions
in $^{87}$Rb and $^{85}$Rb respectively (in ground electronic
state 5S$_{1/2}$). For Raman transition between the hyperfine
states, we utilize far-off resonant excitation through the excited
electronic state 5P$_{3/2}$ (D2 line) near a wavelength of
$\lambda=780.2$~nm. We drive the two Raman transitions with a weak
probe beam, $E_p$, and two strong control lasers, $E_{c1}$ and
$E_{c2}$. The frequency differences between respective beams are
tuned close to hyperfine transition frequencies, $\omega_p -
\omega_{c1} \approx 6.834$~GHz and $\omega_p - \omega_{c2} \approx
3.035$~GHz. The frequency of the probe laser beam, $\omega_p$, is
detuned $\approx$~16~GHz from the D2 line in $^{87}$Rb.

\begin{figure}[th]
\begin{center}
\includegraphics[width=11cm]{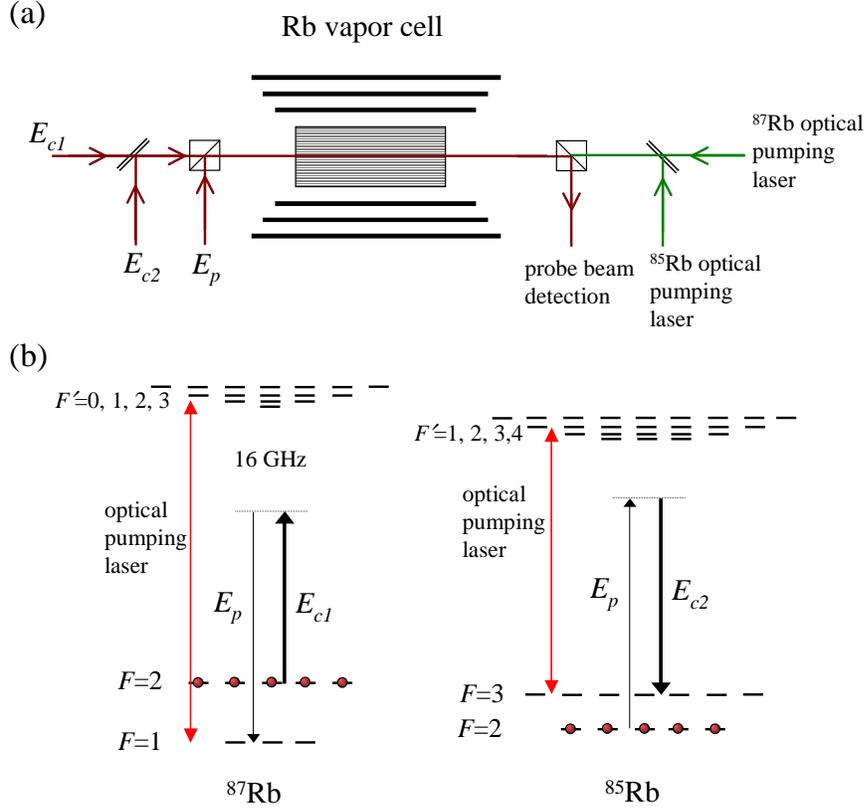}
\end{center}
\caption{ (Color online) Simplified experimental set-up and the
energy level diagram (not to scale) for the two isotopes. The
experiment is performed in a magnetically shielded, natural
abundance Rb vapor cell. Two optical pumping lasers optically pump
the two species to the $F=2$ hyperfine state in the ground
electronic state. With the atoms optically pumped, three
experimental laser beams, $E_p$, $E_{c1}$, and $E_{c2}$ drive two
Raman transitions, one in each isotope. The interference of these
two resonances lead to enhanced refractive index with vanishing
absorption. All three laser beams are far-detuned  from the
single-photon electronic resonance. } \label{experimental_setup}
\end{figure}

We start the experiment by optically pumping both of the atomic
species to the $F=2$ hyperfine state manifold. This is achieved by
two optical pumping lasers locked to $F=1 \rightarrow F'=2$
transition in $^{87}$Rb and $F=3 \rightarrow F'=3$ transition in
$^{85}$Rb respectively. Each optical pumping beam has a power of
about 0.5~W and is obtained by seeding a semiconductor tapered
amplifier with an external cavity diode laser. The optical pumping
beams have a collimated beam radius of $\approx$~3~mm and
counter-propagate the three experimental beams as shown in
Fig.~2(a).

The three experimental beams, $E_p$, $E_{c1}$, and $E_{c2}$, are
derived from a single external cavity master diode laser. The
output of this laser is appropriately shifted by three high
frequency acousto-optic modulators (AOM) in parallel to produce
the desired frequency spacing between respective beams. After the
AOMs, the control beams are amplified by tapered amplifiers to
achieve the required power levels. The frequency of each of the
three laser beams can be tuned by changing the modulation
frequency of the AOMs. This set-up gives us complete control over
the position of the two Raman resonances when we scan the
frequency of the probe laser beam. Further details regarding our
laser system can be found in our previous publications
\cite{YavuzGroup1,YavuzGroup2}. The polarization of the probe beam
is linear and orthogonal to the polarization of the two control
laser beams. The three beams have a collimated beam waist of
$W_0=1.2$~mm at the vapor cell. The probe laser has an optical
power of about 1~mW and is much weaker when compared with the
control lasers ($\approx$~100~mW each).

We run the experiment in a timing cycle where we optically pump
the atoms for about 500~$\mu$s. We then turn-off the optical
pumping beams and turn-on the probe and the control lasers. To
avoid undesired time-dynamics due to sharp edges, we turn-on the
three beams smoothly over about 10~$\mu$s and perform our
measurements at the peak of the pulses. After the beams exit the
vapor cell, we separate the weak probe beam with a high-extinction
polarizer. To determine the gain or the loss on the probe beam, we
measure the intensity of the beam at the peak of its spatial
profile right after it leaves the vapor cell.

\begin{figure}[th]
\begin{center}
\includegraphics[width=7cm]{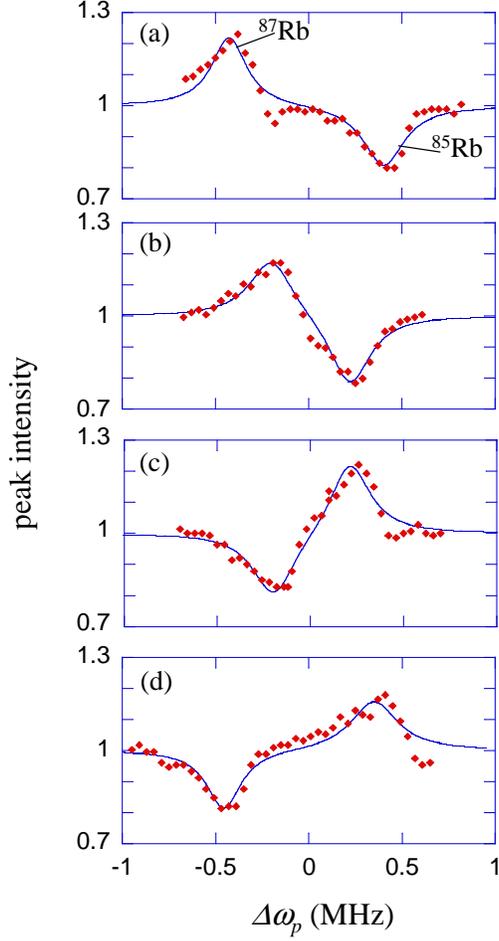}
\end{center}
\caption{ (Color online) The peak intensity of the probe laser
measured after the cell as a function of the frequency of the
probe laser. With the control lasers, we induce a gain resonance
and an absorption resonance on the probe laser. When the probe
laser is resonant with the $F=2 \rightarrow F=1$ Raman transition
in $^{87}$Rb, the beam experiences gain. When it is resonant with
$F=2 \rightarrow F=3$ Raman transition in $^{85}$Rb, the beam
experiences loss. By changing the frequencies of the control
lasers, we can tune each of the Raman resonances. In plots (a),
(b), (c), and (d), the two resonances are spaced by 0.8~MHz,
0.4~MHz, -0.4~MHz, and -0.8~MHz respectively. }
\label{tuning_resonances}
\end{figure}

Figure~3 shows the electromagnetically induced gain and absorption
resonances on the probe beam and our ability to control these
resonances. Here, we measure the peak intensity of the probe beam
as a function of the probe laser frequency. The solid line in each
plot is a fit to the data that assumes each resonance to be a
Lorentzian. When the probe laser is resonant with the $F=2
\rightarrow F=1$ transition in $^{87}$Rb, there is gain on the
beam. When it is resonant with $F=2 \rightarrow F=3$ transition in
$^{85}$Rb, there is loss. By changing the frequencies of the
control lasers, we can control the position of these resonances as
we scan the frequency of the probe laser beam. In Figs.~3(a) and
3(b), the gain resonance happens before the loss resonance whereas
in Figs.~3(c) and 3(d), the situation is reversed.

We proceed with a discussion of the measurement of the refractive
index. Due to the spatial profile of the control lasers, the
refractive index enhancement is larger at the center of the probe
beam when compared with the wings of the beam. As a result, the
probe beam acquires a spatially dependent phase as it propagates
through the atomic medium. In essence, the medium acts as a thin
lens causing the probe beam to focus or defocus after the cell
\cite{YavuzGroup2}. To measure the electromagnetically induced
focusing effect, we measure the transmission of the probe beam
through a 150~$\mu$m wide pinhole placed 2.5~m away from the cell.
If the beam focuses (de-focuses), the beam size decreases
(increases) and the transmission through the pinhole increases
(decreases).

\begin{figure}[th]
\begin{center}
\includegraphics[width=10cm]{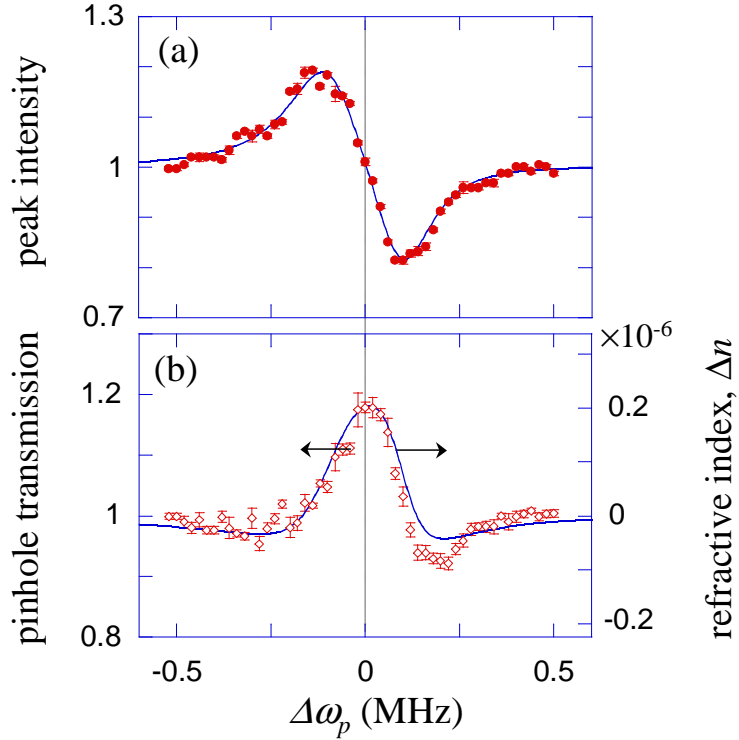}
\end{center}
\caption{ (Color online) Experimental demonstration of the
refractive index enhancement at the point of vanishing absorption.
(a) The peak intensity of the probe beam and (b) the transmission
through a pinhole as a function of probe laser frequency. The
transmission through the pinhole changes as a result of focusing
or defocusing of the beam due to spatial dependence of the
refractive index. The pinhole transmission and therefore the
refractive index is maximized at the point of vanishing
absorption. The solid line in (a) is a fit that assumes the two
resonances to be Lorentzian. The solid line in (b) is the
calculated refractive index change at the peak of the spatial
profile based on the fit of (a). We see good qualitative agreement
between pinhole transmission data and our calculation. }
\label{index_enhancement}
\end{figure}

In Figure 4, we measure the peak intensity of the probe beam
simultaneously with the transmission through the pinhole. The
pinhole transmission is appropriately normalized to take into
account the gain or absorption of the beam. In essence, the
intensity measurement of Fig.~4(a) probes the imaginary part of
the susceptibility ($\chi''$) whereas the pinhole transmission
measurement of Fig.~4(b) probes the real part of the
susceptibility ($\chi'$). For this data, we adjust the control
laser frequencies such that the two resonances are separated by
0.2~MHz which is roughly twice the width of each resonance. The
pinhole transmission, and therefore the refractive index, is
maximized at the point of vanishing gain or loss on the beam. The
solid line in Fig.~4(a) is a fit to the data that assumes each
resonance to be Lorentzian. The solid line in Fig.~4(b) is the
calculated refractive index based on the fit to the data of
Fig.~4(a). We see good qualitative agreement between the
refractive index calculation and the pinhole transmission data.

Currently, our experiment suffers from an undesired cross-coupling
of the two optical pumping processes. The $^{87}$Rb optical
pumping laser tries to pump the $^{85}$Rb atoms to the wrong state
and vice versa. Due to this cross coupling, our optical pumping
efficiency is low and the observed refractive index in our
experiment ($\Delta n \approx 10^{-6}$) is about an order of
magnitude lower than the ideal limit for our experimental
parameters. This problem can be solved by going to a different
laser system and pumping both species to their lower energy
hyperfine ground states ($F=1$ in $^{87}$Rb and $F=2$ in
$^{85}$Rb) which we plan to implement in the near future.

One key advantage of our scheme is that by increasing the
intensities of the control lasers one can obtain the maximum
possible refractive index of the medium while maintaining
vanishing absorption \cite{Yavuz,Kocharovskaya1}. The maximum
refractive index of an alkali vapor at a density of about
10$^{15}$ atoms/cm$^3$ is $\approx$~2. The collisional broadening
of the excited state prevents the refractive index to be increased
further in an alkali vapor \cite{Kocharovskaya1}. Demonstrating a
refractive index of 2 in an alkali vapor with vanishing absorption
may have significant practical implications. Since the wavelength
of light is reduced inside the vapor, an optical microscope with
two times better resolution may be constructed. Similarly, a
better lithographic resolution may be obtained inside the medium.
Furthermore, our scheme is general and may be applied to molecular
species such as molecular $N_2$. In molecules, Raman transitions
between vibrational and rotational states may be utilized, which
may allow refractive index values approaching 10 to be obtained.

We also note the relation of our work to the growing field of
negative index (left-handed) meta-materials \cite{Pendry}.
Recently, there has been a number of theoretical suggestions that
achieve negative refractive index with low loss in the optical
region of the spectrum in atomic systems
\cite{Oktel,MandelNIMinAtoms,FleischhauerNIM}. We note that, by
changing the order of gain and absorption resonances [such as in
Fig.~\ref{tuning_resonances}(c) and
Fig.~\ref{tuning_resonances}(d)], our scheme may also be used to
obtain a reduced refractive index ($n<1$). At large densities,
this effect may be used to obtain a negative value of the
dielectric permittivity, $\epsilon=1+\chi'<0$, with vanishing
absorption. While this is not sufficient to construct a negative
index medium (since one also needs negative permeability,
$\mu<0$), it may be possible to combine our technique with a
strong magnetic resonance in another specie and obtain a negative
refractive index. A detailed analysis of this effect will be among
our future investigations.

We thank Thad Walker and Mark Saffman for helpful discussions and
Dan Sikes and J. P. Sheehan for their assistance with the
experiment. This work was supported by start-up funds from the
University of Wisconsin at Madison.


\begin{references}

\bibitem{ScullyBook} M. O. Scully and M. S. Zubairy, {\it Quantum Optics} (Cambridge
University Press, Cambridge, 1997).

\bibitem{Scully1} M. O. Scully, Phys. Rev Lett. {\bf 67}, 1855 (1991).

\bibitem{Scully2} M. O. Scully and M. Fleischhauer, Phys. Rev. Lett. {\bf 69},
1360 (1992).

\bibitem{Fleischhauer1}  M. Fleischhauer, C. H. Keitel, M. O. Scully, C. Su, B. T.
Ulrich, and S. Y. Zhu, Phys. Rev. A {\bf 46}, 1468 (1992).

\bibitem{Rathe}  U. Rathe, M. Fleischhauer, S. Y. Zhu, T. W. Hansch, and M. O.
Scully, Phys. Rev. A {\bf 47}, 4994 (1993).

\bibitem{Keitel} M. Macovei and C. H. Keitel, J. Phys. B: At. Mol.
Opt. Phys. {\bf 38}, L315 (2005).

\bibitem{Zibrov} A. S. Zibrov {\it et al.},  Phys. Rev. Lett.
{\bf 76}, 3935 (1996).

\bibitem{Yavuz} D. D. Yavuz, Phys. Rev. Lett. {\bf 95}, 223601 (2005).

\bibitem{Kocharovskaya1} P. Anisimov and O. Kocharovskaya, invited talk, 38th Winter
Colloquium on Physics of Quantum Electronics, Snowbird, UT (2008).

\bibitem{Harris1} S. E. Harris, Phys. Today {\bf 50}, No. 7, 36 (1997).

\bibitem{Kocharovskaya2} O. Kocharovskaya and P. Mandel, Phys. Rev. A {\bf 42}, 523
(1990).

\bibitem{Harris2} A. Kasapi, M. Jain, G. Y. Yin, and S. E. Harris, Phys. Rev.
Lett. {\bf 74}, 2447 (1995).

\bibitem{Hau1}  L. V. Hau, S. E. Harris, Z. Dutton, and C. H. Behroozi,
Nature {\bf 397}, 594 (1999).

\bibitem{Kash} M. M. Kash {\it et al.},
Phys. Rev. Lett. {\bf 82}, 5229 (1999).

\bibitem{Fleischhauer2} M. Fleischhauer and M. D. Lukin, Phys. Rev. Lett. {\bf 84},
5094 (2000).

\bibitem{Phillips} D. F. Phillips {\it et. al.}, Phys. Rev. Lett. {\bf 86}, 783 (2001).

\bibitem{Hau2} C. Liu, Z. Dutton, C. H. Behroozi, and L. V. Hau, Nature {\bf
409}, 6819 (2001).

\bibitem{Wang} J. L. Wang, A. Kuzmich, and A. Dogariu, Nature {\bf 406}, 277
(2000).

\bibitem{Boyd} G. M. Gehring {\it et al.}, Science {\bf 312}, 895 (2007).

\bibitem{Imamoglu} H. Schmidt and A. Imamoglu, Opt. Lett. {\bf 21}, 1936 (1996).

\bibitem{Harris3} S. E. Harris and Y. Yamamoto, Phys. Rev. Lett. {\bf 81}, 3611
(1998).

\bibitem{Zhu} H. Kang and Y. Zhu, Phys. Rev. Lett. {\bf 91}, 093601 (2003).

\bibitem{YavuzGroup1} B. E. Unks, N. A. Proite, and D. D. Yavuz, Rev. Sci. Instrum.
{\bf 78}, 083108 (2007).

\bibitem{YavuzGroup2} N. A. Proite, B. E. Unks, J. T. Green, and D. D. Yavuz, Phys.
Rev. A. {\bf 77}, 023819 (2008).

\bibitem{Pendry} J. B. Pendry, Phys. Rev. Lett. {\bf 85}, 3966 (2000).

\bibitem{Oktel} M. O. Oktel and O. E. Mustecaplioglu,
Phys. Rev. A {\bf 70}, 053806 (2004).

\bibitem{MandelNIMinAtoms} Q. Thommen and P. Mandel, Phys. Rev. Lett. {\bf 96}, 053601 (2006).

\bibitem{FleischhauerNIM} J. Kastel, M. Fleischhauer, S. F. Yelin,
and R. L. Walsworth, Phys. Rev. Lett. {\bf 99}, 073602 (2007).


\end{references}
\end{document}